\documentclass[11pt,letterpaper,oneside]{article} 
\usepackage{setspace, graphicx, fullpage, fancyhdr, amssymb, amsmath, epsfig, array, multirow, tabularx, lscape, booktabs, sidecap, subfig, longtable, enumitem}

\usepackage{color,soul}

\usepackage{caption}
\usepackage[super]{nth}

\usepackage[flushleft]{threeparttable}

\usepackage[pagebackref=true]{hyperref}

\usepackage[section]{placeins}

\usepackage{booktabs}
\usepackage[table,xcdraw]{xcolor}

\usepackage[comma]{natbib}

\setcitestyle{aysep={}} 
\setcitestyle{citesep={;}}

\usepackage[utf8]{inputenc} 

\usepackage[margin=1in]{geometry}

\usepackage[font=small]{caption} 

\hypersetup{
	unicode=false,          
	pdftoolbar=true,        
	pdfmenubar=true,        
	pdffitwindow=false,     
	pdfstartview={FitH},    
	pdftitle={},    
	pdfauthor={},     
	pdfsubject={Subject},   
	pdfcreator={Creator},   
	pdfproducer={Producer}, 
	pdfkeywords={keywords}, 
	pdfnewwindow=true,      
	colorlinks=true,       
	linkcolor=blue,          
	linkbordercolor={0 0 0},  
	citebordercolor={0 0 0},
	citecolor=blue,        
	filecolor=black,      
	urlcolor=blue,         
}

\begin{document}
		\pagestyle{plain} 
	\title{\Large\textbf{Networks of Power: Analyzing World Leaders' Interactions on Social Media}}
	\author{\textbf{Evgeniia Iakhnis} (\href{mailto:iakhnis@usc.edu }{iakhnis@usc.edu }),
	\textbf{Adam Badawy}
	(\href{mailto:abadawy@usc.edu}{abadawy@usc.edu})} 
	\maketitle 
	
	\begin{abstract} 

World leaders have been increasingly using social media platforms as a tool for political communication. However, despite the growing research on governmental accounts on social media, virtually nothing is known about interactions \textit{among} world leaders. Using a novel, cross-national dataset of Twitter communication for leaders of 193 countries for the period of 2012-2017, we construct retweet and mention networks to explore the patterns of leaders’ communication. We use social network analysis to conclude that the leaders' interactions on social media closely resemble their interactions in the offline world. Besides, consistent with the democratic peace theory that underscores a special connection between democracies, we identify political regime as the main predictor of clustering between countries on Twitter. Finally, we explore the patterns of the leaders' centrality to identify features that determine which leaders occupy more central positions in the network. Our findings yield new insights on how social media is used by government actors, and have important implications for our understanding of the impact of new technologies on the new forms of diplomacy.

\vspace{0.5cm}

\vspace{1cm}

\end{abstract}

	\singlespacing
	

	\newpage 
    
    \doublespacing
    
    \section{Introduction}
    
    With the ever-growing influence of social media on our day-to-day lives, it is not surprising that world leaders have been increasingly using social media platforms as a  tool for political communication. To be specific, these days an overwhelming 90\% of governments have active social media accounts\footnote{\url{http://twiplomacy.com/blog/twiplomacy-study-2017/}}.  Unsurprisingly, such drastic proliferation of governmental accounts on social media propelled a new strand of academic research focused on leaders' digital communication and diplomacy \citep{barbera2017new, munger2018elites, zeitzoff2018does}. From this new research agenda we know that political leaders use social media strategically to divert attention from domestic problems, bolster regime legitimacy, and suppress opposition \citep{barbera2018from, gunitsky2015corrupting, pearce2015democratizing}. 
    
    Despite these recent advances, the important limitation of the existing literature is that it only investigates how political leaders communicate with \textit{the public}. However, besides talking to the public, leaders also interact with \textit{each other}. For example, during the 2014 Crimean Crisis the Russian Ministry of Foreign Affairs and the US Sate Department engaged in a Twitter hashtag war.\footnote{\url{https://www.washingtonpost.com/news/worldviews/wp/2014/04/25/russia-hijacks-u-s-state-departments-ukraine-hashtag}} Most prominently, since his election in November 2016, the US President Donald Trump brought Twitter diplomacy to a new level by actively engaging with foreign leaders and announcing key foreign policy decisions online.\footnote{\url{http://thehill.com/policy/defense/336659-trumps-diplomacy-by-twitter-sets-off-firestorm}} With the number of leaders and governments represented online growing steadily, the leaders seem to be embracing a new form of diplomacy that bypasses the traditional offline diplomatic channels. However, possibly due to the data limitations, until now there has been no systematic research exploring interactions \textit{between} leaders on social media platforms. 
    
    Thus, the primary goal of this paper is to provide a first look into the world leaders' interactions on social media. How and how much do leaders interact with each other on social media? Which leaders are most interconnected, and why? And which leaders play the central role in the global social media network? To answer these questions, we are using a novel dataset of all social media communication between leaders of 193 U.N. member countries and match it with key political, geographical, and sociodemographic variables. We find that leaders' interactions on social media closely resemble their interactions in the offline world with leaders communicating mostly within their geographic regions or along the similar level in political hierarchy.  Besides, consistent with a large body of literature in international relations, we find that regime type plays an important role in the way Twitter communities are formed. Finally, we explore the patterns of centrality within the leaders' network and find that democratic leaders tend to occupy more central positions in the network, possibly due to the more transparent and multi-directional communicative tendencies they adopt.

    \section{How do Leaders Interact on Social Media?}

In this paper, we examine how leaders strategically interact with each other on social media platforms, such as Twitter. The strategic usage of mass media by political leaders has been long explored by the scholars of political communication \citep{scheufele2006framing,walgrave2006contingency}. More recently, scholars started adopting the existing theories of political communication to the internet domain and investigate how political leaders use social media. While most studies explore leaders' online behavior within particular county context \citep{munger2018elites, zeitzoff2018does,golbeck2010twitter, grant2010digital, enli2013personalized}, a few others adopt a large-n approach \citep{barbera2017new, bulovsky2018authoritarian, barbera2018from}. Existing research indicates that political leaders adopt social media strategically in response to domestic unrest and actively use it to divert attention from domestic problems \citep{barbera2017new}. Rather than simply censoring, leaders have learned to use online platforms to their advantage; for example, \cite{gunitsky2015corrupting} describes four mechanisms that authoritarian regimes use to co-opt social media: counter-mobilization, discourse
framing, preference divulgence, and elite coordination. Similarly, \cite{pearce2015democratizing} shows how authoritarian regime in Azerbaijan uses social media to harass opposition online. 

Despite the growing research on the way leaders and governments use social media, the overwhelming majority of studies have investigated how political leaders communicate with the public. With the number of leaders and governments represented online growing every year, there has been no systematic research exploring interactions \textit{between} leaders on social media platforms. 

Why is online communication between leaders important? First, it represents a new form of diplomacy that bypasses traditional channels and unfolds openly in front of domestic and international audiences. This new form of diplomacy -- ``twitplomacy'' -- remains largely unexplored and under-theorized \citep{su2015twitplomacy, strauss2015digital}. Second, online interactions between leaders might shed light on important conflictual or collaborative relationships that emerge between world leaders offline. It might provide valuable insights into the leaders' respective agendas, strategies, and bargaining processes. For example, \cite{zeitzoff2018does} shows that the leaders of Hamas and Israeli IDF engaged in a true ``Twitter war'' that brought the conflict to an unprecedented level of publicity. Direct Twitter interactions were used to shape the conflict narrative, demonstrate resolve, attract supporters abroad, and influence public opinion, thus, having direct impact on the conflict.

When exploring world leaders' interactions on Twitter, we first need to establish the most common mode of these interactions. First, we can expect that, compared to the ordinary Twitter users, political leaders would engage with each other at lower rates than ordinary Twitter users. It has been found that political leaders mostly use social media platforms as a top-down channel to broadcast information \citep{barbera2017new}. According to \cite{groll2015}, the fact that leaders rarely respond to or interact with their follows shows that ``world powers have embraced Twitter more as a propaganda tool than as a two-way method of communication." If leaders are not highly engaging with their constituents, then we can expect even lower levels of engagement with other leaders. Second, and perhaps more importantly, communication between political leaders represents a public act visible to all followers of an actor. Friendly or aggressive engagement with a leader of a foreign country might have unexpected reaction among the followers; for example, Donald Trump's aggressive mentioning of Theresa May on Twitter sent shock waves in both US and UK \citep{smith2017}. With the whole country watching, mentioning or retweeting other leaders on Twitter might represent a costly act that might backfire or bring negative publicity. Thus, holding everything else constant, we would expect that communication between leaders is highly strategic and occurs more rarely than between ordinary users.

Following this thesis of strategic communication, retweeting should be especially rare among the world leaders as they would rarely choose to display words of some foreign leader to their constituents. Mentioning should be a more preferred way of communication as it establishes some kind of a connection with a foreign leader/government without losing domestic legitimacy \citep{strauss2015digital}. Following this logic, we hypothesize:\\

\textit{H1: Method of communication: Mentioning will be a more preferred way of communication between leaders than retweeting.} \\

Considering that mentioning and retweeting represents a costly act for a leader that might backfire or result in a loss of legitimacy, investigating the patterns of such engagement can reveal some telling information about the leaders' usage of social media platforms. The key question here is: When leaders do retweet or mention, \textit{whom} do they engage with? One possible answer is that leaders' interactions on social media closely resemble their interactions in the offline world. If this is the case, we should see leaders mentioning/retweeting leaders that they interact most diplomatically within certain geographic regions or along the similar level in political hierarchy. In other words, we would see leaders forming mention/retweet communities by regions or type of actor (i.e. Ministers of Foreign Affairs would form communities with other Ministers of Foreign Affairs). Following this logic,\\
   
   \textit{H2: Clustering in network: Leaders' interactions on social media would resemble their interactions in the offline world with nodes clustering by region and type of actor.}\\
   
   While it seems plausible that leaders would cluster by region and type of actor, we also expect that the regime type might play a role in the way Twitter communities are formed. Specifically, we expect that leaders of democratic countries might cluster together replicating their special relationship on the international arena. A large body of literature in International Relations indicates that democratic countries form a special community based on shared values and norms \citep{risse1995cooperation,maoz1993normative,dixon1994democracy}. In his seminal ``Cooperation among Democracies," \cite{risse1995cooperation}
argues that democracies have a strong collective identity based on shared values
such as human right, the rule of law, and democratic governance. Considering strong patterns of cooperation and engagement between democratic leaders and governments in real life, we could expect to detect these patterns of engagements in the online sphere as well:\\

 \textit{H3: Interactions among Democracies: Leaders from Democratic states are more likely to  engage with other democratic leaders.}\\


Finally, as any other network, the world leaders' network should contain a center and a periphery \citep{csermely2013structure,barbera2015critical}. Traditional communication theory states that a minority of users called influentials drive trends on behalf of the majority of ordinary people \citep{rogers1962diffusion}. These individuals could be described as hubs, connectors, opinion leaders, or simply the most informed, respected, and well-connected members of a certain network or society \citep{cha2010measuring,katz1955personal}. Influence is frequently measured through the number of a user's retweets, as it indicates the ability
of that user to generate content with pass-along value, and a number of mentions, as it indicates the ability of that user to engage others in a conversation \citep{cha2010measuring}. Following the existing theories of communication, we expect that certain leaders will occupy central positions in the network and receive a lion share of interactions from other users; while leaders on the periphery might be barely engaged with at all. Not surprisingly, the the top influentials in ordinary Twitter networks are mostly celebrities and public figures \citep{cha2010measuring}; however, what can explain popularity patterns among the world leaders?

One possible explanation is that online hierarchies might closely replicate international hierarchies we observe in the real world \citep{donnelly2006sovereign,lake2013legitimating}. As some countries are more powerful and dominant than others, online networks could closely reflect this offline pecking order. While international hierarchies are not simply based on material capabilities \citep{nye2004soft,hall1997moral}, empirical analysis indicates that economic strength still plays a substantial role in the countries' international status. For example, according to \cite{renshon2016status}, the correlation between material capabilities and status typically hovers between 0.5 and 0.75. Therefore, more dominant countries, as measured by their economic strength, might attract more engagement online than leaders who represent countries with limited material resources. Following this logic, we hypothesize that:\\

 \textit{H4: Centrality: Leaders of high income countries will occupy more central positions in the leaders' network.}\\
 
 Another possibility is that democratic leaders occupy more central positions within social networks due to the differences in online behavior between democratic and autocratic politicians/institutions. While autocratic leaders might be equally active on social media, they adopt a so-called uni-directional communication style that ``involves the projection of opinions with little to no interaction in the other direction'' \citep[4]{bulovsky2018authoritarian}. Conversely, democratic leaders engage in multi-directional communication that involves open and circular flow of discourse with engagement of different viewpoints. According to \cite{bulovsky2018authoritarian}, such differences in style depends on the regime's incentive structure. Authoritarian power structure incentivizes leaders to use social media accounts for national or international self-promotion rather than engaging in conversation. In democracies, free and fair elections push leaders to be more transparent, accountable, and responsive \citep{schmitter1991democracy}, that results in
 ``consistent media presence that exhibits multi-directional communicative tendencies'' \citep[4]{bulovsky2018authoritarian}. Thus, due to multi-directional nature of democratic leaders' social media accounts we can expect that:\\
 
  \textit{H5: Centrality: Leaders of democratic countries will occupy more central positions in the leaders' network.}\\

    \section {Data}
    
    \subsection{Twitter Dataset}
\begin{figure}[!h] 
	\begin{center}
	\caption{Volume of unique users and tweets by month}
	\includegraphics[width = 5in]{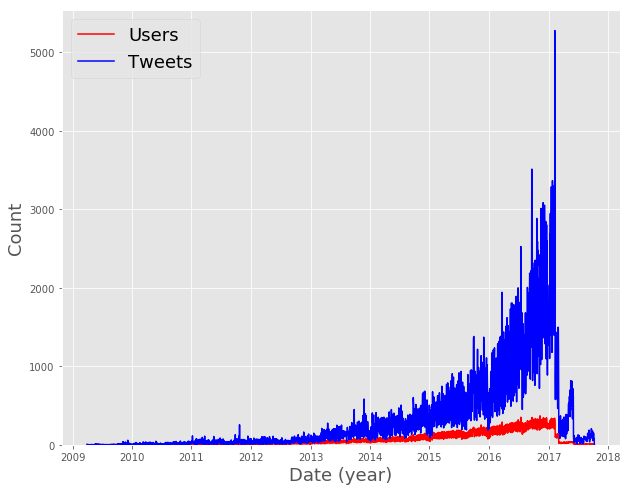}
		\label{fig:volume}
	\end{center}
\end{figure}

    In order to test these theoretical predictions, we use a dataset of all Twitter communication between world leaders (heads of state, heads of government, and ministers of foreign affairs) of 193 U.N. member countries. For each country, a list of relevant names and institutions was identified using the publicly available United Nations Protocol and Liaison Service website (\url{www.un.it/protocol}) as of August 2016. Every name and institution on the list was matched with a respective Twitter account if it exists and has been active (contains at least ten or more posts). Both personal and institutional accounts in local and foreign languages were collected, while parody or fake accounts were carefully excluded.\footnote{We assume that in most cases social media messages are posted by the
leader’s communication staff. However, the specific person posting from the account is irrelevant for our analysis since we are interested in general communication strategy rather than leaders' personal preferences or attitudes.} Using Twitter's Rest API, all Twitter communication from these accounts was gathered in a large dataset of 878,241 tweets that contains all Twitter communication that the leaders engaged in from January 1, 2012 to June 1, 2016 or during their tenure (for further information about the dataset see \cite{barbera2017new, barbera2018from}).   The Figure \ref{fig:volume} shows the time line of the volume of the tweets and the users who produced these tweets during the aforementioned period.\footnote{The shape of the graph reflects the data collection procedure: the further we move from the August 2016 timestamp, the fewer accounts exist in our dataset as we move beyond most leaders' tenure. Such 'decay' occurs only for personal accounts.}
    
Our theoretical argument suggests that the leaders' interactions on social media will be predicted by their regime type, geographical region\footnote{We use the World Bank classification of countries into regions:
East Asia and Pacific (15\%), Europe and Central Asia (27\%), Latin America and
Caribbean (17\%), Middle East and North Africa (11
South Asia (1\%), and Sub-Saharan Africa (24\%).}, and level of economic development. We use the revised Polity IV scores \citep{marshall2002polity} to classify countries as either autocratic, semi-autocratic, semi-democratic, or democratic.\footnote{We use the conventional cut-offs for Polity2, whereby: (autocracy: -10:-8), (semi-autocracy: -7:0), (semi-democracy:1-7), (democracy: 8-10).} Besides, to measure the levels of economic and population development, we use GDP and rates of internet penetration from the International Telecommunication Union (ITU).

    Table \ref{tb:descriptive} reports some aggregate statistics of the dataset. Our dataset covers the communication of 570 leaders who created over 1.1 million tweets during the data collection period (or their tenure time), with the majority of them being original posts. Table \ref{tb:distribution} reports the distribution of accounts in our dataset by type of actor, regime type, and region. As we can see, the accounts are distributed quite evenly among heads of governments, heads of states, and ministers of foreign affairs. The distribution of accounts by regions also mostly corresponds to the number of countries in these regions. As for the middle part of Table \ref{tb:distribution}, it indicates that our dataset over-represents democratic and semi-democratic countries and under-represents autocratic states.\footnote{Some of the obvious missing cases are China, North Korea, and Syria that simply do not maintain governmental Twitter accounts.}

    \begin{table}[!h]
\centering
\caption{Twitter Data Descriptive Statistics}
\label{tb:descriptive}
\begin{tabular}{@{}ll@{}}
\toprule
\multicolumn{1}{c}{\textbf{Statistic}} & \multicolumn{1}{c}{\textbf{Count}} \\ \midrule
\# of Leaders in Dataset               & 570                                \\
\# of Tweets                           & 1,177,497                          \\
\# of Retweets                         & 308,324                            \\ \bottomrule
\end{tabular}
\end{table}

\begin{table}[!h]
\centering
\caption{Distribution of Accounts by Type of Actor, Regime, and Region}
\label{tb:distribution}
\begin{tabular}{|ll|ll|ll|}
\hline
\textbf{Type of Actor} & \textbf{Count} & \textbf{Regime} & \textbf{Count} & \textbf{Region}              & \textbf{Count} \\ \hline
Head of Government     & 180            & Autocracy       & 20             & Sub-Saharan Africa                       & 106            \\
Head of State          & 192            & Semi-autocracy  & 67             & South Asia                         & 24            \\
MFA                    & 198            & Semi-democracy  & 123            & Middle East \& North Africa & 66             \\
                       &                & Democracy       & 234            & Latin America \& Caribbean                       & 99            \\
                       &                &                 &                & North America                & 11             \\
                       &                &                 &                & Europe \& Central Asia                      & 205             \\
                       &                &                 &                & East Asia \& Pacific                & 53             \\ \hline
\end{tabular}
\end{table}

\subsection{Retweet and Mention Networks}

In order to explore the leaders' communication patterns, we constructed two networks corresponding to two main types of interaction on Twitter: retweet and mention networks. Retweet network contains nodes with a direct link between them if one user retweeted a post of another. In contrast, a direct link on a mention network indicates that one user mentioned another in his or her post. While leaders retweet and mention a wide variety of different users including international organizations, media outlets, and even ordinary citizens, here we are only interested in communication among heads of state, heads of government, and ministers of foreign affairs. Thus, we limit our analysis to the networks consisting exclusively of leaders' accounts, with other nodes removed. Moreover, considering our interest in the patterns of international rather than domestic social media communication, we only keep the links that connect leaders with leaders of \textit{other} countries on the list.\footnote{It is widely common for leaders to mention/retweet other leaders of the same country. For example, ministers of foreign affairs often retweet heads of state, and vice versa.}

Table \ref{tb:networks} shows the descriptive statistics of the retweet and mention network. The number of nodes indicates the number of unique leaders in each network. The numbers are somewhat lower than the overall number of leaders in the dataset (570) as any leaders who did not mention/retweet or were mentioned/retweeted during the period of interest were excluded. The number of edges indicates the number of unique interactions between leaders. Maximum weighted in-degree indicates the highest number of times any leader in the network was mentioned/retweeted during the period of data collection. Conversely, maximum weighted out-degree shows the highest number of times any leader in the network mentioned/retweeted other leaders during the time period. Finally, density is the number of observed edges divided by all possible number of edges between leaders.

As we can see, both networks described in the table are sparse networks; however, the mention network is roughly twice more dense than the retweet network. Besides, it contains significantly higher number of nodes and edges. Such differences in the leaders' retweet and mention networks indicate that mentioning is a more preferred way of communication between leaders than retweeting. Compared to ordinary Twitter users, world leaders refrain from retweeting each other, possibly, to preserve legitimacy among their constituents. This descriptive result supports our Hypothesis 1 that mentioning will be a more preferred way of communication between leaders than retweeting.

\begin{table}[]
\centering
\caption{Descriptive statistics of the Leaders' Retweet and Mention Network.}
\label{tb:networks}
\begin{tabular}{@{}l|ll@{}}
\multicolumn{1}{c|}{\textbf{}} & \multicolumn{1}{c}{\textbf{Retweet Network}} & \multicolumn{1}{c}{\textbf{Mentions Network}} \\ \midrule
\# of nodes                    & 447                                          & 500                                           \\
\# of edges                    & 3,550                                        & 7,162                                        \\
Max weighted in-degree                  & 340                                          & 780                                        \\
Max weighted out-degree                 & 6                                          & 46                                           \\
Density                        & 0.018                                       & 0.029                                       
\end{tabular}
\end{table}

    \section {Results}
    
Considering our descriptive finding that mentioning is a more preferred way of communication between leaders than retweeting, we focus our further investigation of leaders' communication patterns on the mentions network. For the sake of space, the respective results for the retweet network could be found in the Online Appendix. In this section we first explore the features that explain the leaders' clustering in the network. Do leaders form online communities resembling their offline interactions? Second, we investigate  how exactly leaders cluster \textit{within} features. For example, if leaders form online geographic communities, which regions cluster together? Finally, we  explore centrality patterns in the leaders' network. In other words, which leaders are the most important in the mention network, and why?

    

Before we start answering these questions, it might be useful to explore when and for what purpose world leaders actually mention each other. Qualitative analysis of a random sample of tweets containing mentions indicates that leaders predominantly mention other leaders in the following situations:

\begin{enumerate}
    \item During state visits, summits, or other diplomatic events;
    \item Congratulating and sending best wishes on national holidays;
    \item Expressing condolences on tragic events, such as acts of terrorism or natural disasters;
    \item Personal birthday wishes or farewells to foreign colleagues who step down from their respective positions;
    \item Expressing support for a policy decision made by a foreign leader;
    \item Expressing criticism for some foreign policy act or a policy decision.
\end{enumerate}

The last category is the most rare but also, perhaps, the most interesting. For example, we see Presidential Administration of Ukraine criticizing Prime Minister of Russia for their policies in Crimea or Minister of Foreign Affairs of Turkey attacking MFA of Sweden for an improper and inaccurate statement. Public mentioning of another country on Twitter in such critical context allows leaders to directly communicate with the citizens of that foreign country and broader international audiences. Appendix 1 provides further examples of tweets that illustrate these different themes, along with the wordcloud that displays the most common words used in the mentions tweets. 

\begin{figure}[!h] 
	\begin{center}
	\caption{Leaders' Mention Network}
	\includegraphics[width = 5.5in]{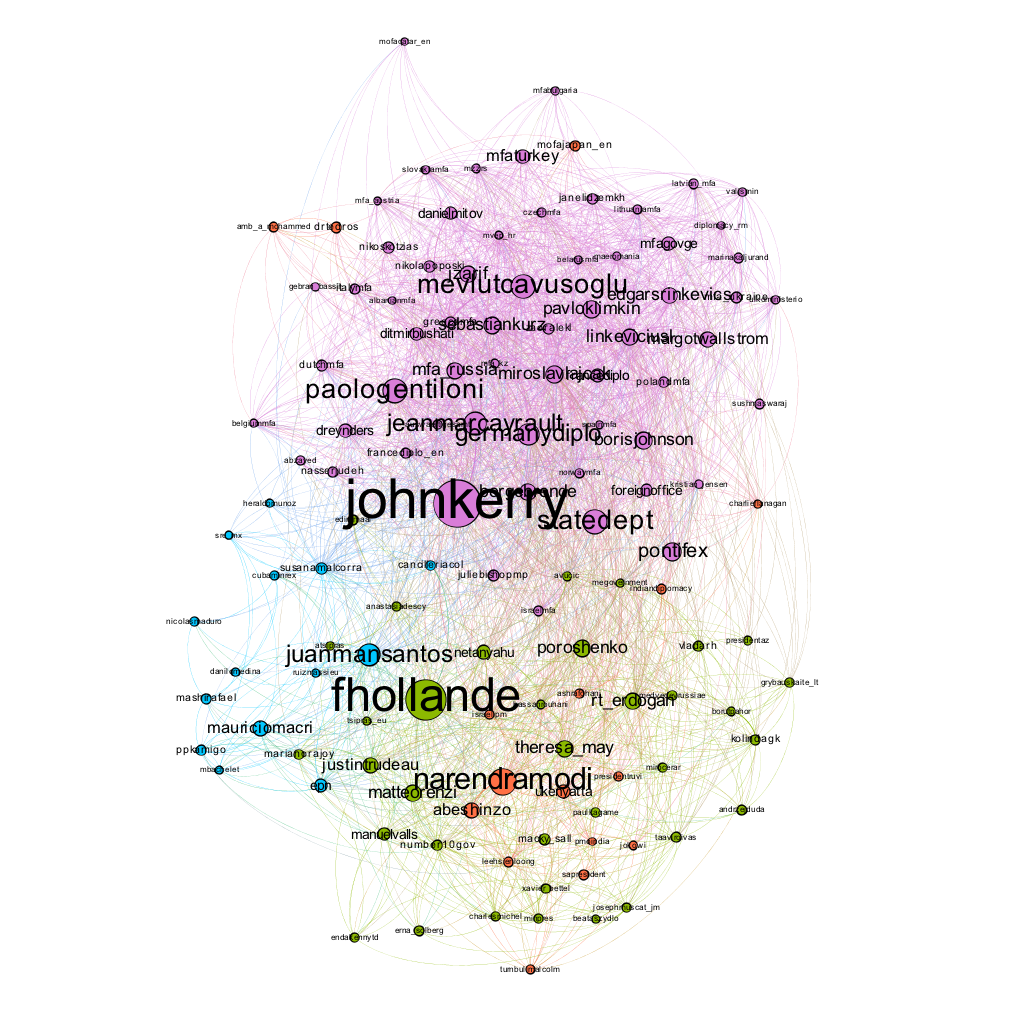}
		\label{fig:gephi}
	\end{center}
\end{figure}

While qualitative investigation of individual tweets is undoubtedly important, the analysis of the whole network is needed in order to understand general patterns of interaction between world leaders. We begin our analysis with visual examination of the world leaders' mention network. Figure \ref{fig:gephi} displays links between leaders with node sizes representing the weighted indegree (how many times a leader was mentioned by other leaders).\footnote{To make the figure more readable, we eliminated the nodes with weighted indegree below 20.} We can see, for instance, that two leaders -- John Kerry and François Hollande -- have the highest indegree in our network. The colors correspond to network communities constructed by the Louvain method. The Louvain algorithm seeks to find the best partitions by maximizing the so-called modularity score, which is a measure usually used to evaluate the quality of partitions being produced by community detection algorithms \citep{de2011generalized}.\footnote{The modularity of a partition is a value between [-1,1] that measures the difference between the density of a network and the expected number of links between nodes if the network was randomly configured \citep{barabasi2013,newman2004,newman2006}.} Overall, Figure \ref{fig:gephi} shows that certain nodes cluster together; however, it does not indicate what might explain such clustering.

  \subsection{Leaders' clustering and interaction patterns}

 Clustering between nodes of similar types (homophily) is one of the most important aspects of networks. Figure \ref{fig:gephi} shows that nodes cluster together based on their connections to other nodes; however, in this paper we do not only care to see nodes' clustering based on their connections, but we aim to explore how certain node features explain this clustering. In order to see which features explain the clustering in the given network, assortativity coefficient has been used \citep{newman2018networks,newman2002assortative}.  Assortativity refers to the extent to which similar nodes are connected in a given network. Assortativity coefficient is typically quantified by computing the Pearson correlation coefficient of the degrees at each end of links in the network if the feature in question is numerical and by using the modulairty measure if the feature is categorical \citep{newman2002assortative}. Assortativity coefficient ranges from [-1,1] for both categorical and numerical features, with higher values indicating higher assortativity. For example, if many edges of a network connect nodes with a certain feature (for example, many edges connect leaders from a particular geographic region), this attribute or feature will have a high assortativity coefficient. 
  
  \begin{table}[!h]
\centering
\caption{Mention Network Assortativity Coefficients}
\label{tab:assort}
\begin{tabular}{@{}lc@{}}
\toprule
\textbf{Feature}  & \textbf{Assortativity Coefficient} \\ \midrule
Type of actor     & 0.41                               \\
Geographic region & 0.27                               \\
Regime            & 0.019                             \\
Population        & -0.01                              \\
Income level      & 0.05                               \\
Internet access   & 0.1                               \\ \bottomrule
\end{tabular}
\end{table}

Assortativity coefficient has been widely used to determine which features explain clustering of users on social media, such as Twitter. For example, \cite{bliss2012twitter} use assortativity coefficient to show that levels of happiness explain clustering of the users in a massive Twitter network. In our case, we expect that leaders' interactions on Twitter resemble their interactions online, thus, the nodes should cluster by region and type of actor (Hypothesis 2). Table \ref{tab:assort} displays a list of attributes of the leaders' accounts and respective assortativity coefficients. As we can see, a type of actor, that indicates whether an account belongs to a Head of State, a Head of Government, or a Minister of Foreign Affairs, has the highest coefficient meaning that actors of the same position tend to cluster together more. Similarly, the leaders from same geographic regions tend to form strong online communities as predicted in our hypothesis. The other features, such as regime type, level of economic development, and internet access do not seem to play an essential role in the assortativity of the world leaders' network. Assortativity coefficients table for the retweet network presented in the Appendix shows a similar pattern with type of actor and geographic region explaining clustering patterns.



\subsection{Heatmaps}

Assortativity coefficients presented in the previous section indicate that world leaders do not seem to cluster by regime type; however, that does not necessarily mean that the regime type does not play any role in the way Twitter communities are formed. Assortativity coefficient for regime type reflects clustering along \textit{all categories} of the attribute, including autocracies and semi-autocracies that might not interact much with each other. In order to properly test our third hypothesis that democratic countries form a special online community on Twitter, we need to look \textit{within} features by breaking them down in particular categories. 

\begin{figure}[!h] 
	\begin{center}
	\caption{Clustering by Specific Regions}
	\includegraphics[width = 5in]{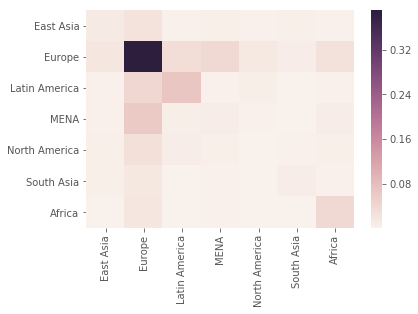}
		\label{fig:heatmap_region}
	\end{center}
\end{figure}

To illustrate the concept, let us first look at the heatmap in Figure \ref{fig:heatmap_region} that breaks down the geographic clustering into specific regions. Each box in the heatmap represents a count probability of edges, where one node has an attribute from the X-axis and another node has an attribute from the Y-axis \citep{newman2018networks}. For example, looking at the heatmap of clustering by region, we can see that the edges where both nodes are European leaders have the highest count probability of 0.4 (in other words, they constitute 40\% of the total number of edges in the entire mention network). Besides, the heatmap shows that the count probability of edges between European and Middle Eastern leaders is also quite high at around 0.063 and 0.067 between Latin American Leaders. Overall, rather than simply concluding that leaders on Twitter cluster by region, we can say specifically whether the leaders of particular regions tend to interact the most. 

\begin{figure}[!h] 
	\begin{center}
	\caption{Clustering by Specific Type of Regime}
	\includegraphics[width = 5in]{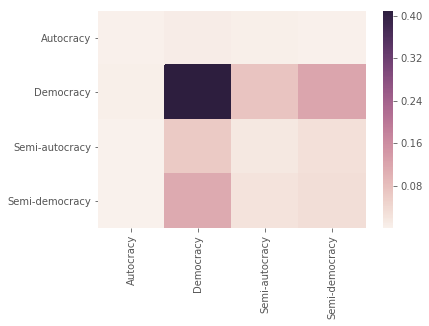}
		\label{fig:heatmap_regime}
	\end{center}
\end{figure}

Moving on to Hypothesis 3, heatmap in Figure \ref{fig:heatmap_regime} breaks down the clustering by regime into specific types of government. The heatmap clearly shows that leaders of democratic countries do cluster together on Twitter. The dark box in the middle of the heatmap indicates that the edges where both nodes are democratic leaders have a very high count probability of 0.4. We can also note the moderately high count probability of edges connecting democratic and semi-democratic leaders (around 0.12). At the same time, almost white boxes of edges between autocratic and autocratic/semi-autocratic leaders indicate extremely low levels of online interaction between these countries. Therefore, we can conclude that the heatmap in Figure \ref{fig:heatmap_regime} provides support for our hypothesis that democratic 
leaders are more likely to engage with other democratic leaders on Twitter. Similarly to the offline world, democratic countries form a special community on this social media platform.

\subsection{Predictors of Social Media Centrality}

Our final hypotheses 4 and 5 concern the patterns of leaders' centrality on social networks. We expect that leaders of high income and democratic countries will occupy more central positions in the network. In order to test this relationship, we must first select the most appropriate operationalization of ``centrality'' as measurement of importance of a node in a network. 

The importance of a node in a network is one of the most important questions in network science. Considering a vast variety of applied problems (importance of an individual in a network of people; a bank in a financial transactions network; a web-page on the internet; a country in a trade network), multiple centrality measures suitable for various problems have been developed. Generally there are are two classes of centrality measures. The first group of centrality measures captures how many connections a node has and includes such measures as: degree centrality (how many connections a node has and has two versions for directed networks: in-degree and out-degree); Eigenvector centrality (sum of the centralities of a node's neighbors, suitable for undirected networks); and PageRank centrality (centralities of a node's neighbors divided by their out-degrees, suitable for directed networks). The second group of centrality measures capture the position of a node in a network. The most notable measure from this group is betweenness centrality that calculates a number of shortest paths between a pair of nodes that passes through the node in question. 

In this paper we use weighted in-degree centrality to capture importance of world leaders. In-degree centrality belongs to the first group of centrality measures. While position centrality (second type) is important, we do not believe that it measures the kind of importance that we are aiming to capture here. For example, states or state leaders who serve as interlocutors between two regions of the world or different political campuses, would have high betweenness centrality, but they usually are not important states in terms of militaristic, economic, and cultural/soft power. Indegree indicates that we are mostly interested in \textit{incoming} connections that occur when a leader in question is being mentioned or retweeted by other leaders. We choose indegree over more sophisticated measures like PageRank, as the latter usually give too much importance to few nodes while giving very small importance values to the rest of the nodes, making the comparison of importance for most nodes quite difficult.\footnote{Additionally, the random-walker model here is not necessary, since we believe that a world leader mention is a costly signal (unlike retweets in a typical retweet network in most Twitter studies) that signals the importance of the person being mentioned.} Finally, we use \textit{weighted} indegree as it captures the number of times a person was mentioned rather than a mere fact of mentioning, which is an important indicator of the importance of a node or in our case.

\begin{figure}[!h] 
	\begin{center}
	\caption{Distribution of Weighted Indegree}
	\includegraphics[width = 4in]{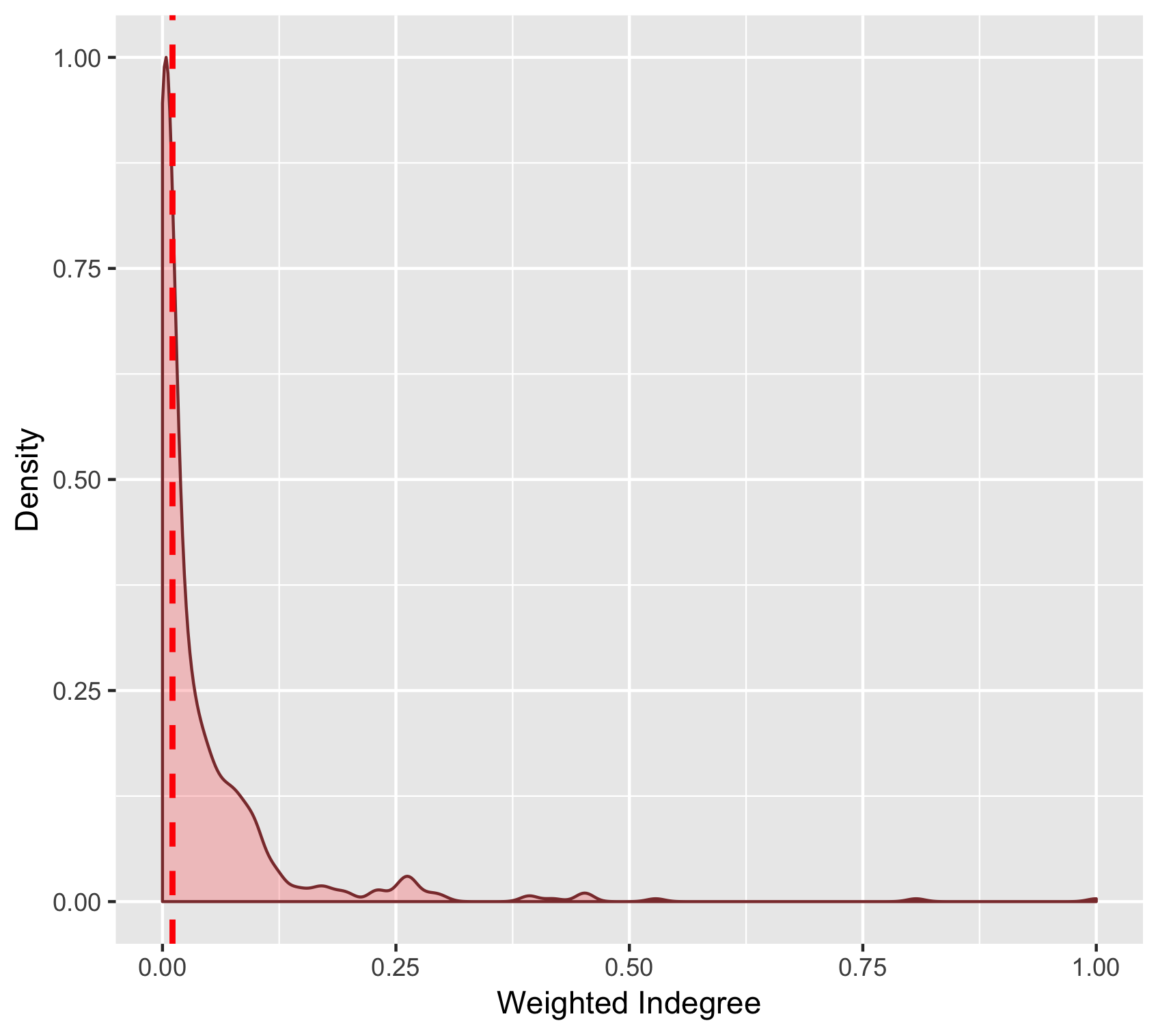}
		\label{fig:indegree_den}
					\footnotesize \caption* {\textit{Note}: Figure shows distribution of weighted indegree. Dotted line indicates the median.}
	\end{center}
\end{figure}

The distribution of weighted indegree is presented in Figure \ref{fig:indegree_den}. A heavy-tailed distribution (also called power-law or scalefree) with a few accounts having a high level of indegree has been found typical for the distribution of the number of ties of a person in a social network \citep{barabasi1999emergence, caldarelli2007scale}. Often such heterogeneous levels of activities follow the well-known and widely applicable law postulated by Pareto, which states that 80\% of the effects are induced by 20\% of the causes \citep{muchnik2013origins}. The Pareto law seems to apply remarkably well to the weighted indegree distribution in our network: 20\% of the top accounts produce 75\% of mentioning activity. We define these 20\% of top accounts as ``central" and the rest 80\% as ``peripheral", and use logistic regression to explore which factors predict centrality of an account.

Table \ref{tab:logit} displays the coefficient estimates for a set of logistic regressions of social media centrality on income level and regime type. The first model introduces a set of control variables that pertain to the type and quality of the account rather than the characteristics of a country. Account age measures how old the Twitter account is (\# of days since a user signed up). Controlling for this variable is important as older accounts would quite logically accumulate more mentions or retweets over time. Statuses count measures how many tweets a user has produced (includes original tweets and retweets). Burstiness is is a measure of burstiness of a system or phenomenon (also called the coefficient of variation) that is measured by subtracting the mean of a distribution from its standard deviation then dividing it by the sum of the mean and the standard deviation. This metric is bounded from [-1,1] where 1 means it is a bursty signal; 0 means neutral; and -1 means regular (periodic) signal.\footnote{For more on this metric, read \cite{goh2008burstiness}.} Own language is a dichotomous variable coded as 1 if whether an account is maintained in a country's local language and 0 for English. Actor is a categorical variable indicating a leader's position: Head of State, Head of Government, or Minister of Foreign Affairs (MFA). Type of account is a dichotomous variable coded as 1 for personal account and 0 for institutional account.

\begin{table}[!h] \centering 
  \caption{Mention Network Centrality} 
  \label{tab:logit} 
  \begin{threeparttable}
\begin{tabular}{@{\extracolsep{5pt}}lcccc} 
\\[-1.8ex]\hline \\[-1.8ex] 
\\[-1.8ex] & \multicolumn{4}{c}{Network Centrality (Top 20\%)} \\ 
\\[-1.8ex] & (1) & (2) & (3) & (4)\\ 
\hline \\[-1.8ex] 
 Account age & 4.393$^{**}$ & 3.566$^{**}$ & 3.900$^{**}$ & 3.205$^{**}$ \\ 
  & (0.685) & (0.715) & (0.705) & (0.738) \\ 
  Statuses count & 1.973 & 3.139$^{*}$ & 2.045 & 2.809$^{*}$ \\ 
  & (1.016) & (1.275) & (1.051) & (1.313) \\ 
  Burstiness & $-$0.388 & 0.002 & 0.296 & 0.503 \\ 
  & (2.548) & (2.769) & (2.479) & (2.662) \\ 
  Own language & $-$0.917$^{**}$ & $-$0.824$^{**}$ & $-$0.897$^{**}$ & $-$0.880$^{**}$ \\ 
  & (0.271) & (0.285) & (0.282) & (0.295) \\ 
  Actor: Head of State & 0.370 & 0.573 & 0.324 & 0.570 \\ 
  & (0.355) & (0.378) & (0.371) & (0.390) \\ 
  Actor: MFA & 1.374$^{**}$ & 1.479$^{**}$ & 1.354$^{**}$ & 1.502$^{**}$ \\ 
  & (0.339) & (0.358) & (0.352) & (0.370) \\ 
  Personal account & 2.191$^{**}$ & 2.471$^{**}$ & 2.295$^{**}$ & 2.589$^{**}$ \\ 
  & (0.295) & (0.327) & (0.311) & (0.342) \\ 
  Middle income &  & 0.139 &  & 0.149 \\ 
  &  & (0.468) &  & (0.466) \\ 
  High income &  & 0.755 &  & 0.633 \\ 
  &  & (0.641) &  & (0.637) \\ 
  Internet users &  & 0.978 &  & 1.171 \\ 
  &  & (0.976) &  & (0.971) \\ 
  Democracy Dummy &  &  & 1.077$^{**}$ & 0.879$^{**}$ \\ 
  &  &  & (0.314) & (0.337) \\ 
  Constant & $-$4.946$^{**}$ & $-$6.090$^{**}$ & $-$5.895$^{**}$ & $-$6.916$^{**}$ \\ 
  & (1.828) & (2.047) & (1.808) & (1.996) \\ 
 N & 531 & 491 & 478 & 459 \\  
\hline \\[-1.8ex] 
\multicolumn{5}{l}{$^{*}$p $<$ .05; $^{**}$p $<$ .01} \\ 
\end{tabular}
\begin{tablenotes}
			\small
			\item \textit{Notes}: Table displays logit coefficients, standard errors in parentheses. All non-dichotomous measures have been rescaled from 0 to 1.
		\end{tablenotes}
	\end{threeparttable}
\end{table}

Model 1 presents a baseline that explains variation in centrality solely through the type and quality of account. Unsurprisingly, the age of account is strongly correlated with online centrality - the older the account, the more mentions it accumulates. The level of overall online activity also matters: the more tweets a user produces, the more likely he or she is to be central in the mention network. Ministers of Foreign Affairs tend to occupy more central positions in the network compared to other types of accounts, possibly, due to their stronger international and diplomatic activity. Posting in a country's local language is negatively associated with high network centrality as such accounts are primarily maintained for domestic consumption. Finally, personal accounts tend to be more central than institutional accounts, perhaps, due to their more engaging nature. 

Now, moving to our variables of interest, Models 2 and 3 test hypotheses 4 and 5 respectively. Model 2 looks at the effect of income level measured through GDP and rates of internet penetration. GDP is fitted as a categorical variable with three levels corresponding to the terciles of the distribution (``Low income'' , ``Middle income'', ``High income'').\footnote{In Appendix, we replicate the analysis using the continuous measure and show that the results stay same.} Insignificant coefficients in Model 2 indicate that high income countries are not more likely to occupy central positions in the leaders' network than countries with low or middle income. Rates of internet penetration do not seem to be a key predictor of network centrality. H4 finds no support. Model 3 tests the effect of regime type using the
dichotomous specification of regime type in which countries with a Polity IV score of 6 or higher are coded as democracies \citep{jaggers1995tracking, barbera2017new}. The significant and positive coefficient of the dummy indicates that democratic leaders are indeed more likely to be central in the network than autocratic leaders.\footnote{In  Appendix,  we  replicate  the  analysis  using  a  continuous variable of Polity IV and electoral democracy index from the Varieties of Democracies Project (V-Dem). The results stay same regardless of the specification.}

\begin{figure}[!h] 
	\begin{center}
	\caption{Effect of Regime on Probability of Network Centrality}
	\includegraphics[width = 5.5in]{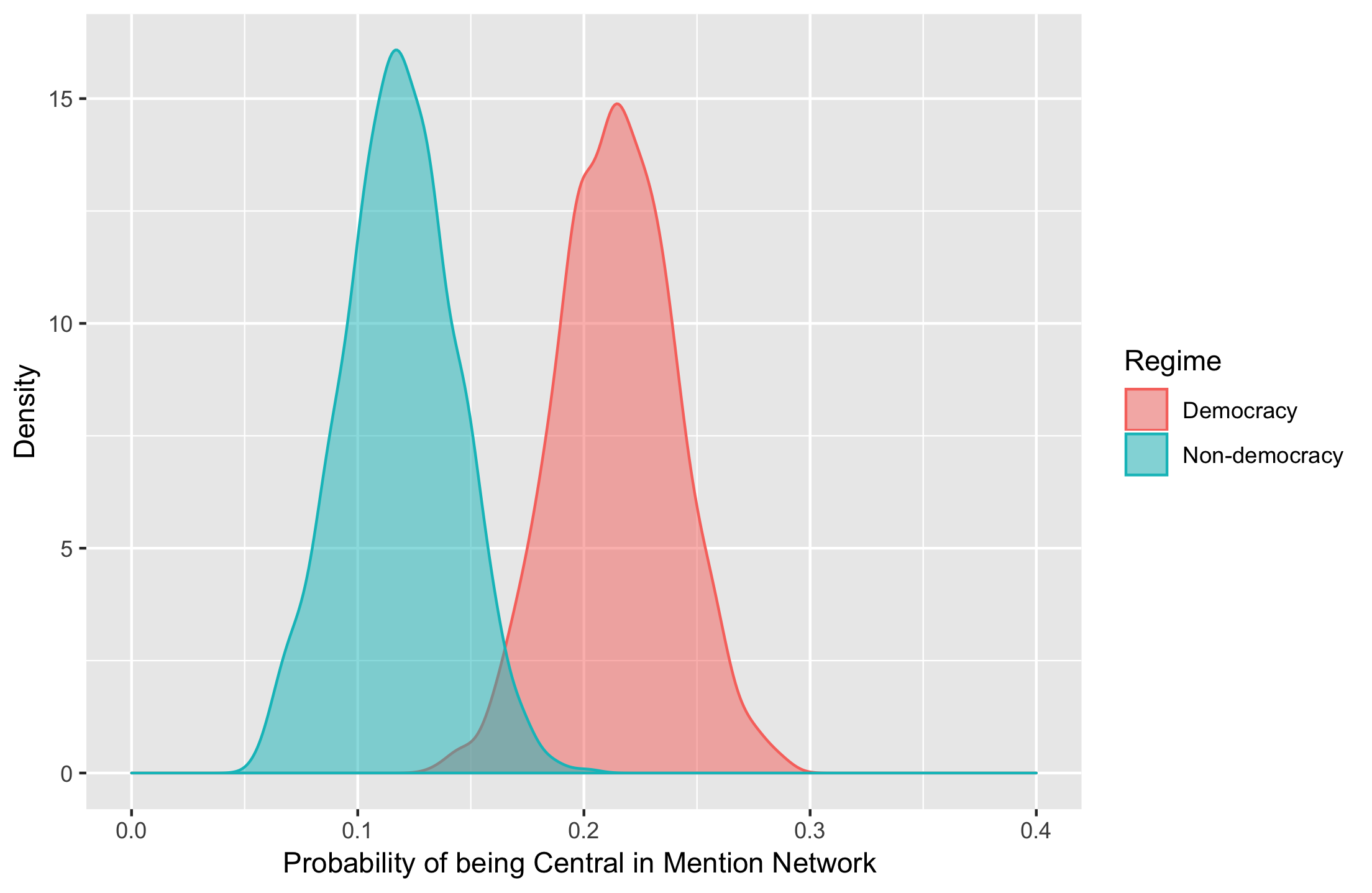}
		\label{fig:indegree_plot}
					\footnotesize \caption* {\textit{Note}: Figure displays predicted probability distributions derived from 1,000 bootstraps from logistic regression models of the impact of regime type on the probability of being central in mention network, controlling for other significant variable in Model 4 of Table \ref{tab:logit}.}
	\end{center}
\end{figure}

The result is robust to adding income variables as shown in the saturated Model 4. Considering that economic development might be a strong predictor of democracy \citep{cheibub1996makes}, it is useful to control for income to better isolate the relationship between democracy and centrality \citep{bulovsky2018authoritarian}. Figure  \ref{fig:indegree_plot} shows predicted probabilities of being central in mention network for democratic and non-democratic leaders based on the saturated model.\footnote{These predicted probabilities are derived from 1,000 bootstrapped logit models of the impact of regime on network centrality, keeping control variables at their means and modes.} The predicted probability of being central in the network is 9.6\% higher for democratic leaders than non-democratic leaders (p $<$ 0.01 from a bootstrapped difference of means test). The figure strongly suggests that democratic leaders occupy more central position within the leaders' mention network as we suggested in Hypothesis 5. 

\section{Checking Network Stability Over Time}

In this paper we treat the mention network as a static network and ignore its dynamic nature. Such approach is common in network analysis \citep{goyal2018dyngraph2vec}; however, some additional validation is necessary to make sure that such static approach is justified. If a network changes significantly across time, treating it as static  can be misleading since the analysis would only reflect the nature of the network at the end of the data collection, but not at various temporal points prior to that. To eliminate this possibility, we test the stability of our mention network by examining the weighted pagerank and the normalized weighted in-degree distributions across time. 

\begin{figure}
    \caption{PageRank (left) and Weighted in-degree (normalized by sum of in-degree per time period) (right) statistics across the whole time period}
   \includegraphics[width=0.5\textwidth]{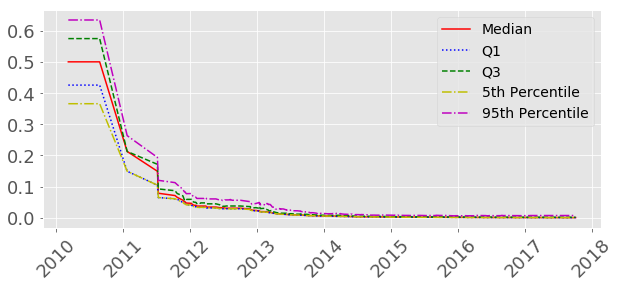}
   \includegraphics[width=0.5\textwidth]{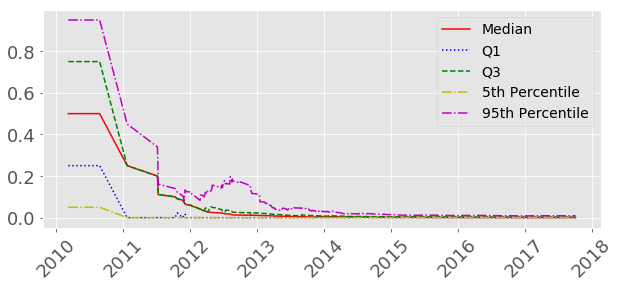}
    \label{fig:valid_whole}
\end{figure}

\begin{figure}
    \caption{PageRank (left) and Weighted in-degree (normalized by sum of in-degree per time period) (right) statistics form 2015 onwards, where more than 86\% of the mention data is represented}
      \includegraphics[width=0.5\textwidth]{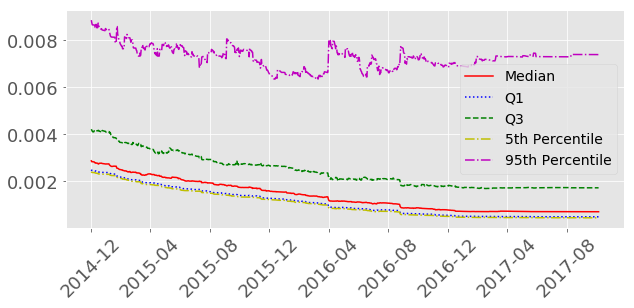}
   \includegraphics[width=0.5\textwidth]{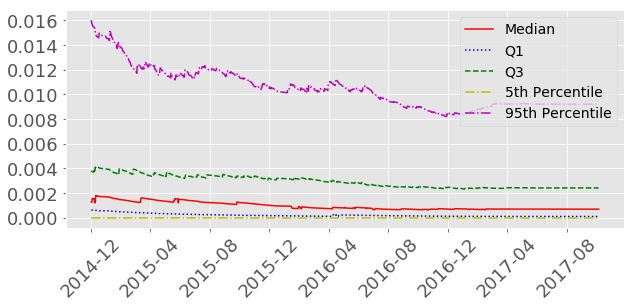}
   \label{fig:valid_2015}
\end{figure}

The way the networks are constructed in figures \ref{fig:valid_whole} and \ref{fig:valid_2015} are by including all the nodes, edges, and weights (number of mentions by node $i$ to node $j$ in a directed network) up to time $t$ with monthly intervals.  In other words, every temporal network includes all the nodes, edges, and weights that occurred in preceding months. By observing weighted pageranks and in-degree distributions, we can see whether the distributions change over time or stay stable. 

In figure \ref{fig:valid_whole} we show five different metrics: \nth{5} percentile, \nth{1} quantile, median, \nth{3} quantile, and \nth{95} percentile. The reason for showing different metrics for each temporal network is that both pagerank and in-degree distributions for twitter data tend to be power-law distributions \citep{barabasi1999emergence}. As mentioned above, a power law distribution means that most users have very low pagerank or in-degree, while some top users have extremely high pagerank or in-degree. In the context of a mention network, this means that few leaders receive the most mentions while most leaders do not get mentioned or get mentioned only a few times. Thus, when assessing a network metric distribution, it is imperative not to look only at the mean or median, but at the top percentiles as well.

Figure \ref{fig:valid_whole} shows that our mention network is very stable from 2013, which is not that surprising, since prior to 2013 there was little data collected (see Figure \ref{fig:volume} for volume of tweets by time). Figure \ref{fig:valid_2015} shows the same metrics, but starting from 2015. Since 86\% of the mention data occurred after this date, it is important to take a deeper look at the metrics trends within this time frame. All the metrics are fairly stable, although the \nth{95} percentile looks slightly less stable as the other metrics. Looking at the difference between the upper and the lower bounds on the y-axis for this metric, we can see that it is quite small in both the pagerank and in-degree distributions. Taken together, figures \ref{fig:valid_whole} and \ref{fig:valid_2015} justify our assumption that the mention network can be treated as a static network, ignoring the temporal variability. 

\section{Discussion and Conclusion}

In this project we use a unique dataset of all Twitter communication between the leaders of U.N. member countries to provide a first look into the world leaders' interactions on social media. We have three key findings. First, we reveal that leaders' interaction on social media closely resemble their interaction in the offline world. Similarly to the offline diplomatic communities, we show that leaders form mention/retweet communities along regional lines and similar levels in political hierarchy. 

Second, we find that the regime type plays a key role in the way Twitter communities are formed. Consistent with the expectations of the democratic peace theory, we find that leaders from democratic states are more likely to engage with other democratic leaders. Finally, we explore the popularity patterns among the world leaders and conclude that leaders of democratic countries tend to occupy more central positions in the network, with leaders of non-democratic countries present at the periphery. One potential explanation of this result is that democratic leaders tend to be more transparent and responsive online \citep{bulovsky2018authoritarian}. Rather than simply using social media for credit claiming and self-promotion, they are wiling to engage in active conversation with their followers. At the same time, we found no indication that online hierarchies might be based on material capabilities. 

While our project provides a first important step towards understanding of the leaders' interactions online, many questions remained unanswered. For example, here we focus on the leaders' network connections, rather than the content of their interactions. Content analysis may provide
greater insight into why and when leaders mention or retweet each other. Are these interactions mostly friendly and diplomatic in nature? Or do leaders sometimes use these tools to publicly criticize or attack their political opponents? These questions should be addressed in future work.

Furthermore, the conclusions made in this paper are unavoidably limited by the scope of the data at hand. First, as mentioned somewhere in the paper, certain leaders and governments are missing from the datatset in a non-random fashion. The dataset under-represents autocratic states as many of them do not have an active Twitter account. Besides, the dataset covers a very limited time period with leaders dropping out when their tenure ends. Thus, the data provides more of a static snapshot rather than a temporal comparison of the leaders' interactions developing and changing overtime. We hope that future data collection efforts will work to solve this issue and provide a more comprehensive investigation of the dynamic nature of the leaders' interactions.

\bibliographystyle{chicago}
\bibliography{wl_network.bib}

\newpage 
\section{Online Appendix}

\subsection{Examples of Mention Tweets}

\begin{figure}[!htbp]
	\centering
	\caption{Examples of Mentions by World Leaders}
	\subfloat[State visits]{\includegraphics[width=0.5\linewidth, height = 6.5cm]{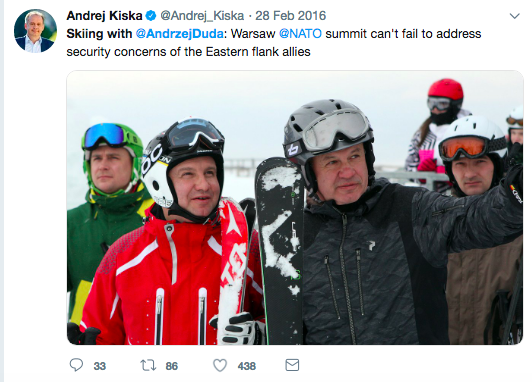}\label{fig5:a}}
	\subfloat[Congratulating on national holidays]{\includegraphics[width=0.5\linewidth, height = 6.5cm]{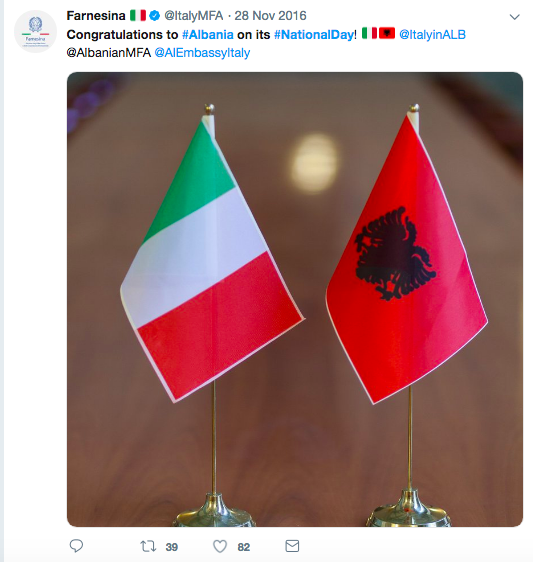}\label{fig5:b}}\\
	\subfloat[Expressing condolences]{\includegraphics[width=0.5\linewidth, height = 6.5cm]{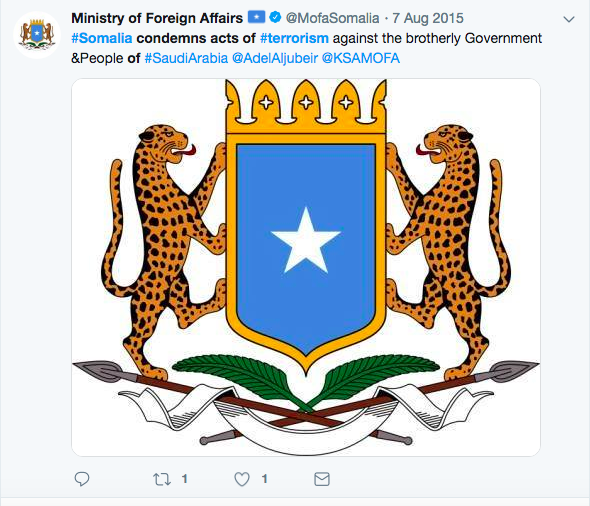}\label{fig5:b}}
	\label{fig:examples1}
\end{figure}

\begin{figure}[!htbp]
	\centering
	\caption{Examples of Mentions by World Leaders}
	\subfloat[Personal messages]{\includegraphics[width=0.5\linewidth, height = 6.5cm]{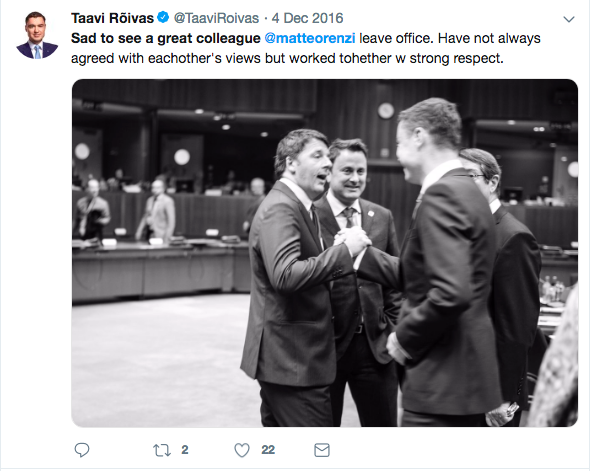}\label{fig5:a}}
	\subfloat[Support for policy decisions]{\includegraphics[width=0.5\linewidth]{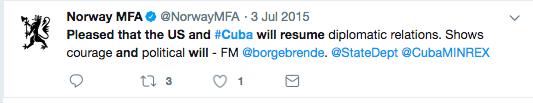}\label{fig5:b}}\\
	\subfloat[Criticism]{\includegraphics[width=0.5\linewidth]{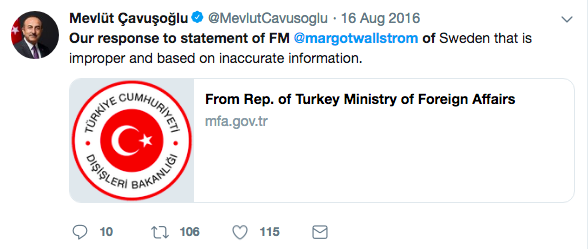}\label{fig5:b}}\\
\subfloat{\includegraphics[width=0.5\linewidth]{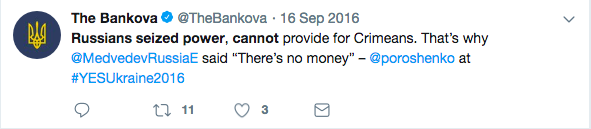}\label{fig5:b}}
	\label{fig:examples2}
\end{figure}

\begin{figure}
\centering
    \caption{Word Cloud of 200 Most Common Words in Mention Tweets}
      \includegraphics[width=0.9\textwidth]{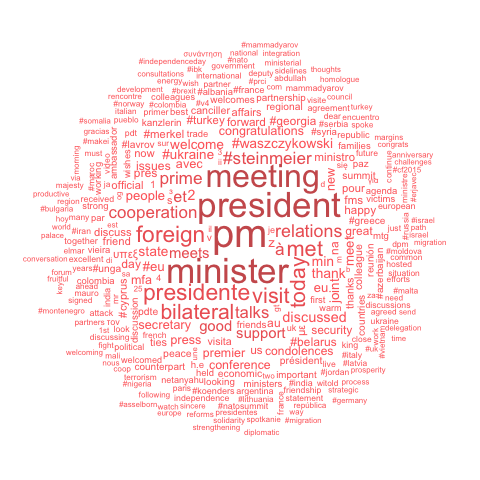}
   \label{fig:wordcloud}
\end{figure}

Figures \ref{fig:examples1} and \ref{fig:examples2} show examples of mention tweets of six categories described in the main text. Figure \ref{fig:wordcloud} shows 200 most common words used when world leaders mention each other. It shows that leaders mostly mention each other in the context of diplomatic visits and bilateral cooperation.

\newpage

\subsection{Robustness Checks}

In this section, we explore the robustness of our main findings. The first model of Table \ref{tab:robust} shows that the alternative measure of income level as log GDP yields similar result: the variable is not statistically significant. Models 2 and 3 probe the robustness of our main finding that democratic leaders have higher levels of network centrality. in Model 2 we use the continuous version of Polity IV's Polity 2 scale. The significant and positive coefficient of the variable indicates that the leaders of more democratic countries are more likely to be central in the mention network. Finally, in Model 3 we use the electoral democracy index as measured by the Varieties of Democracies Project (V-Dem). The advantage of this project from Polity IV and others is that it includes multiples varieties of democracy: electoral, liberal, participatory, deliberative, and egalitarian \citep{lindberg2014v}. Considering that our hypothesis is mostly based on the idea of electoral pressure in democracies, we use the electoral component to check robustness of our main findings. The highly significant and positive coefficient of electoral democracy in Model 3 (the variable ranges from 0 for less democratic to 1 for more democratic) provides confident support for our previous results.

\begin{table}[!h] \centering 
  \caption{Alternative Specifications of the Model} 
  \label{tab:robust} 
  \begin{threeparttable}
\begin{tabular}{@{\extracolsep{5pt}}lccc} 
\\[-1.8ex]\hline \\[-1.8ex] 
\\[-1.8ex] & \multicolumn{3}{c}{Network Centrality (Top 20\%)} \\ 
\\[-1.8ex] & (1) & (2) & (3)\\ 
\hline \\[-1.8ex] 
  Account age & 3.276$^{**}$ & 3.164$^{**}$ & 3.427$^{**}$ \\ 
  & (0.734) & (0.740) & (0.729) \\ 
  Statuses count & 2.554$^{*}$ & 2.828$^{*}$ & 2.837$^{*}$ \\ 
  & (1.284) & (1.325) & (1.310) \\ 
  Burstiness & 0.635 & 0.242 & 0.267 \\ 
  & (2.627) & (2.602) & (2.688) \\ 
  Own language & $-$0.874$^{**}$ & $-$0.856$^{**}$ & $-$0.840$^{**}$ \\ 
  & (0.294) & (0.298) & (0.295) \\ 
  Actor: Head of State & 0.542 & 0.643 & 0.627 \\ 
  & (0.388) & (0.394) & (0.392) \\ 
  Actor: MFA & 1.491$^{**}$ & 1.595$^{**}$ & 1.527$^{**}$ \\ 
  & (0.369) & (0.376) & (0.369) \\ 
  Personal account & 2.551$^{**}$ & 2.652$^{**}$ & 2.585$^{**}$ \\ 
  & (0.338) & (0.349) & (0.341) \\ 
  Internet users & 1.421 & 1.164 & 1.087 \\ 
  & (1.143) & (0.975) & (0.984) \\ 
  Log GDP & 0.146 &  &  \\ 
  & (0.255) &  &  \\ 
  Democracy (dummy)  & 0.903$^{**}$ &  &  \\ 
  & (0.335) &  &  \\ 
  Middle income &  & 0.169 & 0.077 \\ 
  &  & (0.465) & (0.468) \\ 
  High income &  & 0.687 & 0.512 \\ 
  &  & (0.636) & (0.652) \\ 
  Polity IV score &  & 1.767$^{**}$ &  \\ 
  &  & (0.565) &  \\ 
  Electoral democracy &  &  & 2.571$^{**}$ \\ 
  &  &  & (0.845) \\ 
  Constant & $-$6.573$^{**}$ & $-$7.620$^{**}$ & $-$8.251$^{**}$ \\ 
  & (2.203) & (1.991) & (2.115) \\ 
 N & 459 & 459 & 476 \\
\hline \\[-1.8ex] 
\multicolumn{4}{l}{$^{*}$p $<$ .05; $^{**}$p $<$ .01} \\ 
\end{tabular}
\begin{tablenotes}
			\small
			\item \textit{Notes}: Table displays logit coefficients, standard errors in parentheses. All non-dichotomous measures have been rescaled from 0 to 1.
		\end{tablenotes}
	\end{threeparttable}
\end{table}

\newpage

\subsection{Retweet Network results}

In this section, we present our findings for the leaders' retweet network and show that the patterns of communications between leaders on these two networks are remarkably similar. Table \ref{tab:retweet_assort} displays a list of attributes of the leaders' acocunts and respective assortativity coefficients (similarly to Table 4 in the main text). Again, we can see that a type of actor and geographic region have the highest values in the table, indicating that leaders of the same position and from the same geographic regions tend to cluster together more. other features do not seem to play an essential roles in the assortativity of the leaders retweets network.

\begin{table}[]
\centering
\caption{Retweet Network Assortativity Coefficients}
\label{tab:retweet_assort}
\begin{tabular}{@{}lc@{}}
\toprule
\textbf{Feature}  & \textbf{Assortativity Coefficient} \\ \midrule
Type of actor             & 0.38            \\
Geographic region          & 0.38             \\
Regime           & 0.05           \\
Population       & -0.01            \\
Income level    & 0.10             \\
Internet access & 0.16    
  \\ \bottomrule
\end{tabular}
\end{table}

\begin{figure}[h] 
	\begin{center}
	\caption{Retweet Network Type of Government Clustering}
	\includegraphics[width = 5in]{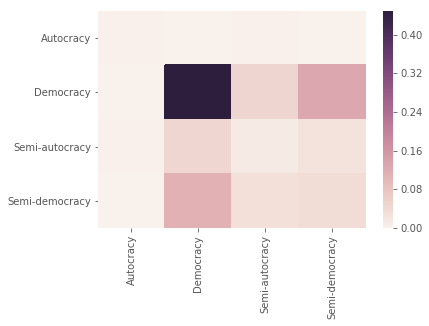}
		\label{retweet_heatmap_regime}
	\end{center}
\end{figure}

\begin{figure}[h] 
	\begin{center}
	\caption{Retweet Network Region Clustering}
	\includegraphics[width = 5in]{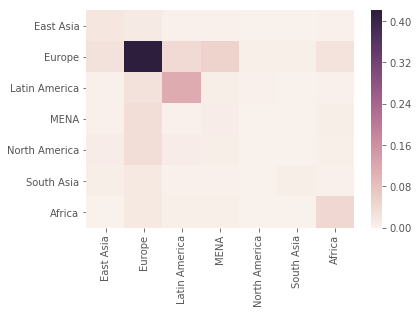}
		\label{retweet_heatmap_region}
	\end{center}
\end{figure}

Moving on to the heatmaps that break down features into specific categories, Figure \ref{retweet_heatmap_region} again shows that edges where both nodes are European leaders have the highest count probability of 0.4.The count probability of edges between Latin American leaders is also quite high at 0.2 meaning that the leaders of Latin American countries tend to retweet each other often. Heatmap in Figure \ref{retweet_heatmap_regime} breaks down the clustering by regime into specific types of government. Similarly for the mention network, it shows that democratic leaders are much more likely to engage with other democratic leaders on Twitter. The dark box in the middle of the heatmap indicates that the edges where both nodes are democratic leaders have a high count probability of 0.4.

Finally, we need to check whether the patterns on leaders' centrality on the retweet network are similar to what we have seen with the mention network. Similarly to our main analysis, we select weighted in-degree centrality as the our measure of centrality. Unsurprisingly, the distribution of retweet weighted indegree is also heavy-tailed with a few accounts having a high level of indegree and most of the accounts having very low levels. Again, we dichotomize the variable following the Pareto law: 20\% of the top accounts produce 80.5\% of all retweeting activity. 

Table \ref{tab:retweet_logit} displays the coefficient estimates for a set of logistic regressions of social media centrality on the similar set of predictors as Table 5 in the main text. Consistent with our main results, Model 2 shows that high income countries are not more likely to occupy central positions in the leaders' retweet network. Rates of internet penetration emerge as a significant predictor in the model; however, the effect is not robust to the addition of the variable of regime (Model 4). Democracy dummy again emerges as a highly significant and robust predictor of leaders centrality on Twitter. 

\begin{table}[!h] \centering 
  \caption{Retweet Network Centrality} 
  \label{tab:retweet_logit} 
  \begin{threeparttable}
\begin{tabular}{@{\extracolsep{5pt}}lcccc} 
\\[-1.8ex]\hline \\[-1.8ex] 
\\[-1.8ex] & \multicolumn{4}{c}{Retweet Network Centrality (Top 20\%)} \\ 
\\[-1.8ex] & (1) & (2) & (3) & (4)\\ 
\hline \\[-1.8ex] 
  Account age & 1.951$^{**}$ & 1.188 & 1.294 & 0.757 \\ 
  & (0.641) & (0.693) & (0.668) & (0.715) \\ 
  Statuses count & 5.428$^{**}$ & 7.538$^{**}$ & 5.483$^{**}$ & 7.000$^{**}$ \\ 
  & (1.146) & (1.616) & (1.163) & (1.606) \\ 
  Burstiness & $-$1.100 & $-$1.455 & $-$0.724 & $-$1.195 \\ 
  & (2.882) & (2.888) & (2.736) & (2.756) \\ 
  Own language & $-$0.953$^{**}$ & $-$0.933$^{**}$ & $-$0.966$^{**}$ & $-$0.933$^{**}$ \\ 
  & (0.262) & (0.276) & (0.273) & (0.280) \\ 
  Actor: Head of State & 0.607 & 0.712 & 0.610 & 0.744 \\ 
  & (0.405) & (0.420) & (0.424) & (0.431) \\ 
  Actor: MFA & 2.204$^{**}$ & 2.246$^{**}$ & 2.214$^{**}$ & 2.286$^{**}$ \\ 
  & (0.350) & (0.363) & (0.367) & (0.375) \\ 
  Personal account & $-$0.008 & 0.208 & 0.011 & 0.205 \\ 
  & (0.266) & (0.282) & (0.278) & (0.291) \\ 
  Middle income &  & $-$0.783 &  & $-$0.709 \\ 
  &  & (0.431) &  & (0.433) \\ 
  High income &  & $-$0.682 &  & $-$0.694 \\ 
  &  & (0.575) &  & (0.577) \\ 
  Internet users &  & 1.807$^{*}$ &  & 1.744 \\ 
  &  & (0.904) &  & (0.923) \\ 
  Democracy (dummy) &  &  & 0.819$^{**}$ & 0.644$^{*}$ \\ 
  &  &  & (0.285) & (0.299) \\ 
  Constant & $-$2.719 & $-$2.944 & $-$3.149 & $-$3.253 \\ 
  & (2.042) & (2.066) & (1.938) & (1.972) \\ 
 N & 527 & 491 & 474 & 459 \\  
\hline \\[-1.8ex] 
\multicolumn{5}{l}{$^{*}$p $<$ .05; $^{**}$p $<$ .01} \\ 
\end{tabular}
\begin{tablenotes}
			\small
			\item \textit{Notes}: Table displays logit coefficients, standard errors in parentheses. All non-dichotomous measures have been rescaled from 0 to 1.
		\end{tablenotes}
	\end{threeparttable}
\end{table}

\end{document}